\documentstyle[preprint,pra,aps]{revtex}
\begin{document}
\draft
\title{\bf Enhancement of  nuclear Schiff moments and time
 reversal violation in atoms due to soft nuclear octupole vibrations}
\author{V.V. Flambaum$^{1}$ and V.G. Zelevinsky$^{2}$}
\address{$^{1}$School of Physics, University of New South Wales,
Sydney 2052, Australia and \\Institute for Nuclear Theory,
University
of Washington, Seattle, WA 98195-1550, USA\\
$^{2}$National Superconducting Cyclotron Laboratory and\\
Department of Physics and Astronomy, Michigan State University,
East Lansing, MI 48824-1321, USA}
\date{\today}
\maketitle

\tightenlines
\begin{abstract}

Parity and time invariance violating (${\cal P},{\cal T}$-odd)
nuclear forces produce ${\cal P},{\cal T}$-odd nuclear moments,
for example, the nuclear Schiff moment. In turn, this moment can
induce electric dipole moments (EDMs) in atoms. We estimate the
contribution to the Schiff moment from the soft collective
octupole vibrations existing in many heavy nuclei. Two possible
mechanisms are considered, mixing of the ground state of an
odd-$A$ nucleus with the octupole phonon state, and ${\cal
P},{\cal T}$-odd admixture to the single-particle wave function of
the valence nucleon. We found practically the same contribution to
the Schiff moment as that of the static octupole deformation
calculated earlier. This confirms a suggestion that the soft
octupole vibrations can replace the (controversial)  static
octupoles in development of the collective Schiff moments. The
values of atomic EDM predicted for $^{223,225}$Ra and $^{223}$Rn
are enhanced by factors up to $10^{3}$ compared to experimentally
studied spherical nuclei $^{199}$Hg and $^{129}$Xe.
\end{abstract}
\vspace{1cm} 

\section{Introduction}

A discovery of the static electric dipole moment (EDM) of an
elementary particle, atom or atomic nucleus would reveal
\cite{ramsey} the simultaneous violation of the invariance with
respect to spatial inversion $({\cal P})$ and time reversal
$({\cal T})$. The best limit on nucleon-nucleon interactions
violating parity and time invariance (${\cal P},{\cal T}$-odd
forces), as well as quark-quark ${\cal P},{\cal T}$-odd
interactions, has been obtained from the measurement of the atomic
EDM in $^{199}$Hg \cite{fortson} calculated in Ref. \cite{FKS86}.
According to the Schiff theorem \cite{ramsey,schiff,sandars}, the
nuclear EDM is screened by atomic electrons. The EDM of an atom
with closed electron subshells is induced by the nuclear Schiff
moment \cite{FKS84} that is defined as a mean square radius of the
dipole charge distribution with the contribution of the
center-of-charge subtracted,
\begin{equation}
{\bf S}=\frac{e}{10}\left[ \langle r^{2}{\bf r}\rangle
-\frac{5}{3Z}\langle r^{2}\rangle \langle {\bf r}\rangle \right].
\label{s}
\end{equation}
Here $\langle r^{n}\rangle \equiv \int \rho ({\bf r})r^{n}d^{3}r$
are the moments of the nuclear charge density operator $\rho$. The
expectation values of the ${\cal P}$-odd vector operators $\langle
{\bf r}r^{2}\rangle$ and $\langle {\bf r}\rangle$ do not vanish
due to the ${\cal P},{\cal T}$-odd contribution to the charge
density while $\langle r^{2}\rangle/Z=r_{{\rm ch}}^{2}$ is the
nuclear mean square charge radius. As a consequence of the
rotational invariance, the existence of the non-zero Schiff moment
requires a non-zero nuclear spin ${\bf I}$,
\begin{equation}
{\bf S}=S~\frac{{\bf I}}{I}.
\label{vect}
\end{equation}

In a search for nuclear structure mechanisms of the enhancement,
it was suggested in Ref.\cite{haxton} that some actinide nuclei
have a level of opposite parity and the same spin close to the
ground state, and this may enhance the level mixing and the
resulting nuclear EDM. In Ref. \cite{FKS84} a similar suggestion
was put forward for the enhancement of the Schiff moment ${\bf
S}$. It was noticed in Ref. \cite{FZ} that, in contrast to ${\cal
P}$-odd ${\cal T}$-even forces, the parity doublets present for a
static pear-shaped deformation can be directly mixed by ${\cal
P},{\cal T}$-odd forces. This mixing might be enhanced because of
close intrinsic structure within the doublet and relatively small
energy splitting. As shown in Refs. \cite{AFS,SAF}, nuclei with
static octupole deformation may have enhanced collective Schiff
moments up to $1000$ times larger than the Schiff moments of
spherical nuclei. An additional enhancement factor appears in this case 
due to the large static intrinsic Schiff moment in the
body-fixed frame which is proportional to the collective octupole moment.

It was suggested in Ref. \cite{hayes} that the soft octupole
vibrations observed in some regions of the nuclear chart may
produce an enhancement similar to that due to static octupole
deformation. This would make heavy atoms containing such nuclei
with large collective Schiff moments attractive for future
experiments in search for ${\cal P,T}$-violation; experiments of
this type are currently under progress or in preparation in
several laboratories. Below we perform the estimate of the Schiff
moment generated in nuclei with the soft octupole mode and show
that the result is nearly the same as in the case of the static
octupole deformation. We consider two simultaneously acting
mechanisms, mixing of the ground state of the odd-$A$ nucleus with
the one-phonon octupole state (``core excitation") and mixing of
the single-particle wave function of the valence unpaired nucleon
with levels of opposite parity through the dynamical octupole
deformation (``particle excitation"). Our analytical estimates are
based on well known general properties of nuclear structure rather
than on specific assumptions concerning the model parameters.

\section{Collective nuclear Schiff moment}

Consider a nucleus with two levels, ground state $|{\rm
g.s.}\rangle$ and excited state $|x\rangle$, of opposite parity
and close energies, $E_{{\rm g.s.}}$ and $E_{x}$, respectively.
Let $W$ be a ${\cal P},{\cal T}$-odd interaction capable of mixing
these unperturbed states. Assuming that the mixing matrix elements
of ${\bf S}$ and $W$ are real, we can write down the Schiff moment
emerging in the actual mixed ground state as
\begin{equation}
{\bf S}=2\frac{\langle {\rm g.s.}|W|x\rangle \langle x|{\bf
S}|{\rm g.s.}\rangle}{E_{{\rm g.s.}}-E_{x}}. \label{wschiff}
\end{equation}
However, as it was explained in Ref. \cite{FKS84}, in the case of
mixing of closely lying single-particle states one should not
expect a large enhancement. For example, in a simple approximate
model, where the strong nuclear potential is proportional to
nuclear density and the spin-orbit interaction is neglected, the
matrix element $\langle {\rm g.s.}|W|x\rangle$ contains the
single-particle momentum operator and is proportional to $(E_{{\rm
g.s.}}-E_{x})$ so that the small energy denominator cancels out.
As mentioned above, the collective Schiff moments in nuclei with
static octupole deformations may be by 2-3 orders of magnitude
stronger than single-particle moments in spherical nuclei.

The mechanism generating the collective Schiff moment is the
following \cite{AFS,SAF}. In the ``frozen" body-fixed frame the
intrinsic collective Schiff moment $S_{\rm intr}$ can exist
without any ${\cal P},{\cal T}$-violation. However, in the
space-fixed laboratory frame the nucleus has a certain angular
momentum rather than orientation, and this makes the expectation
value of the Schiff moment vanish in the case of no ${\cal
P},{\cal T}$-violation. Indeed, the intrinsic Schiff moment is
directed along the nuclear axis, ${\bf S}_{\rm intr}=S_{\rm
intr}{\bf e}$, and in the laboratory frame the only possible
correlation $\langle {\bf e}\rangle \propto {\bf I}$ violates
parity and time reversal invariance. The ${\cal P},{\cal T}$-odd
nuclear forces mix rotational states $|I,\pm\rangle$ of the same
spin and opposite parity and create an average orientation of the
nuclear axis ${\bf e}$ along the nuclear spin ${\bf I}$,
\begin{equation}
\langle e_{z}\rangle =2\alpha \frac{KM}{I(I+1)},
\label{nz}
\end{equation}
where
\begin{equation}
\alpha =\frac{\langle I{-}|W|I{+}\rangle}{E_{+}-E_{-}}  \label{6}
\end{equation}
is the mixing amplitude of the states of opposite parity, $K=|{\bf
I}\cdot {\bf e}|$ is the absolute value of the projection of the
nuclear spin ${\bf I}$ on the nuclear axis, and $M$ is the spin
projection onto the laboratory quantization axis. The observable
Schiff moment in the laboratory frame is then related to the
intrinsic moment as
\begin{equation}
S_{z}=S_{\rm intr}\langle e_{z}\rangle =S_{\rm intr}\frac{2\alpha
KM} {I(I+1)}.
\label{7}
\end{equation}

To estimate the intrinsic Schiff moment, we use the standard
description of the surface of an axially symmetric deformed
nucleus in the body-fixed frame in terms of the multipole
deformation parameters $\beta_{l}$,
\begin{equation}
R(\theta)=R\,[1+\sum_{l=1}\beta_{l}Y_{l0}(\theta)].    \label{rad}
\end{equation}
In order to keep the center-of-mass at the origin we have to fix
$\beta _{1}$ \cite{bm}:
\begin{equation}
\beta_{1}=-3\sqrt{\frac{3}{4\pi}}\sum_{l=2}\frac{(l+1)\beta_{l}\beta_{l+1}}
{\sqrt{(2l+1)(2l+3)}}\,.
\label{beta1}
\end{equation}
We assume that the center of the charge distribution coincides
with the center-of-mass, so that the electric dipole moment
vanishes, $e\langle {\bf r}\rangle=0$, and hence there is no
screening contribution to the Schiff moment (no second term in eq.
(\ref{s})). We also assume a constant density for $r<R(\theta)$.
The intrinsic Schiff moment $S_{\rm intr}$ is then \cite{AFS,SAF}
\begin{equation}
S_{\rm intr}=eZR^{3}\frac{3}{20\pi}\sum_{l=2} \frac{(l+1)\beta
_{l}\beta _{l+1}}{\sqrt{(2l+1)(2l+3)}} \approx \frac{9}{20\pi
\sqrt{35}}\,eZR^{3}\beta _{2}\beta_{3} \ ,
                                                       \label{intrschiff}
\end{equation}
where the major contribution comes from the product
$\beta_{2}\beta_{3}$ of the static quadrupole, $\beta_{2}$, and
octupole, $\beta _{3}$, deformation parameters. For $\beta
_{2}\sim \beta_{3}\sim 0.1$ and $Z=88$ (Ra) we obtain $S_{\rm
intr}\sim 10 ~e \, {\rm fm}^{3}$. The estimate of the Schiff
moment in the laboratory frame [$K=M=I$ in Eq. (\ref{7})]
gives~\cite{SAF}
\begin{equation}
S =S_{\rm intr}\frac{2\alpha I}{I+1} \sim 0.01\,\frac{I}{I+1} ~e
\beta_{2}\beta_{3}^{2}ZA^{2/3} \frac{\eta G}{mr_0 (E_{+}-E_{-})} ,
\label{snum}
\end{equation}
where $\eta G$ is the strength constant of the ${\cal P},{\cal
T}$-odd nuclear potential, traditionally introduced with the aid
of the Fermi weak interaction constant $G$, $R=r_{0}A^{1/3}$, $A$
is the nuclear mass number, and $r_{0}\approx 1.2~{\rm fm}$ is the
internucleon distance.  For the isotope $^{225}$Ra, where
$E_{+}-E_{-}= 55~{\rm keV}$ and $I$=1/2, this analytical estimate
gives the Schiff moment $S \sim 500$ in units $ 10^{-8}\eta e~{\rm
fm}^{3}$; for  $^{223}$Ra ($E_{+}-E_{-}= 50~{\rm keV}$, $I$=3/2)
$S \sim 1000$. The numerical calculation ~\cite{SAF} gives $S$=300
for  $^{225}$Ra, 400 for $^{223}$Ra and 1000 for $^{223}$Rn. To
indicate the accuracy of the calculation we should note that the
difference between the values obtained for the Woods-Saxon
potential (presented above) and Nilsson potential is within a
factor of 2 \cite{comment}.

The values we obtained are several hundred times larger than the
Schiff moment of a spherical nucleus like Hg ($S=-1.4$). An
additional enhancement of the atomic EDM appears due to the
greater nuclear charge in Ra and Rn and close atomic states of
opposite parity \cite{ra1,ra2}. Accurate atomic calculations of
the EDM for atoms of Hg, Xe, Ra, Rn, and Pu have been performed in
Ref. \cite{dzuba}.

${\cal T}$ and ${\cal P}$-odd nuclear forces can also produce
enhanced collective magnetic quadrupole \cite{mag} and octupole
\cite{murray} moments. Although these moments cannot induce the
EDM in the ground state of the closed-shell atoms like Xe, Hg, Rn
and Ra, they can contribute to the EDM of metastable excited
states or of the ground states in open-shell atoms.

Note that the Schiff moment $S$ in Eq. (\ref{snum}) is
proportional to the squared octupole deformation parameter
$\beta_{3}^{2}$. According to \cite{hayes}, in nuclei with a soft
octupole vibration mode the dynamical octupole deformation is of
the order of $\langle \beta _{3}^{2}\rangle \sim (0.1)^{2}$, i.e.
the same as the static octupole deformation in pear-shaped nuclei.
This means that a number of heavy nuclei can have large collective
Schiff moments of dynamical origin. A calculation of these moments
is the main subject of the present paper.

\section{Schiff moment produced by soft octupole vibrations}

A fully microscopic calculation of the Schiff moment in odd-$A$
actinides, even with the aid of simplifying models, such as the
particle-plus-rotor model \cite{leander}, is very complicated,
requires also the solution of the random phase approximation for
the soft octupole mode in adjacent even-even isotopes and contains
many uncertainties in the choice of the parameters. For our
purpose of estimating the magnitude of the effect we use below
simple analytical arguments.

The intrinsic nuclear octupole moment in the body-fixed frame is
given by \cite{bm}
\begin{equation}
O_{\rm intr}= e\int \rho r^3 Y_{30}d^3 r \approx \frac{3}{4\pi}
eZR^3 \beta_{3}.                      \label{introct}
\end{equation}
The intrinsic Schiff moment $S_{\rm intr}$, Eq.
(\ref{intrschiff}), is then
\begin{equation}
S_{\rm intr}=\frac{3}{5\sqrt{35}}O_{\rm intr}\beta_{2}.
\label{schiffoct}
\end{equation}
This relation allows us to approximately express the matrix
elements of the Schiff moment operator in odd-$A$ nuclei with
static axially symmetric quadrupole deformation $\beta_{2}$ in
terms of the matrix elements of the octupole operator which can be
extracted from the observed probabilities of the octupole
transitions in an even-even neighbor. In a similar way, as a
characteristic of the effective dynamic deformation parameter
$\beta_{3}$ in a soft nucleus, it is convenient to use the r.m.s.
value extracted from the reduced octupole transition probability
B(E3) \cite{spear},
\begin{equation}
{\rm B(E3)}_{0\rightarrow 3}=|\langle 1| O_{\rm intr}|0\rangle |^2
\approx\left(\frac{3}{4\pi}eZR^{3}\right)^2\langle\beta_{3}^{{\rm
rms}} \rangle^{2}
\label{b3}
\end{equation}
Here $|1\rangle$ is a low-lying collective (``one-phonon")
octupole excited state.

To find the ${\cal P},{\cal T}$-odd Schiff moment (\ref{wschiff})
we need to know the matrix elements of the ${\cal P},{\cal T}$-odd
nucleon-nucleon interaction. To the first order in the velocities
$p/m$, the ${\cal P},{\cal T}$-odd interaction  can be presented
as \cite{FKS84}
\begin{equation}
\hat{W}_{ab}=\frac{G}{\sqrt{2}}\frac{1}{2m}\left( (\eta_{ab}
\mbox{\boldmath$\sigma$}_{a}-\eta_{ba}\mbox{\boldmath$\sigma$}_{b})\cdot
\mbox{\boldmath$\nabla$}_{a}\delta({\bf r}_{a}-{\bf
r}_{b})+\eta'_{ab}\left[ \mbox{\boldmath$\sigma$}_{a}\times
\mbox{\boldmath$\sigma$}_{b}\right] \cdot \left\{({\bf p}_{a}-{\bf
p}_{b}),\delta({\bf r}_{a}-{\bf r}_{b}) \right\}\right),
\label{wfull}
\end{equation}
where $\{\ ,\ \}$ is an anticommutator, $G$ is the Fermi constant
of the weak interaction, $m$ is the nucleon mass, and
\mbox{\boldmath$\sigma$}$_{a,b}$, ${\bf r}_{a,b}$, and ${\bf
p}_{a,b}$ are the spins, coordinates, and momenta, respectively,
of the interacting nucleons $a$ and $b$. The dimensionless
constants $\eta _{ab}$ and $\eta '_{ab}$ characterize the strength
of the ${\cal P},{\cal T}$-odd nuclear forces; experiments on
measurement of the EDMs are aimed at extracting the values of
these constants.

In the context of our problem one can envisage two efficient
mechanisms generating the observable ${\cal P},{\cal T}$-odd
effects. Firstly, the octupole component of the ${\cal P},{\cal
T}$-odd field produced by the valence nucleon can excite the soft
octupole excitation in the core. For brevity we will call this
mechanism ``core excitation". The second mechanism suggested in
Ref. \cite{hayes}, ``particle excitation", results from the
dynamic octupole deformation of the nuclear mean field that mixes
the opposite parity orbitals of the valence nucleon and leads to a
non-vanishing expectation value of the ${\cal P},{\cal T}$-odd
interaction in the fixed-body frame. As we will show, the value of
the Schiff moment induced by the second mechanism is practically
identical to that obtained earlier in Refs. \cite{AFS,SAF} for the
case of static octupole deformation. We start with the estimate for
the first mechanism.

\subsection{Core excitation}.

Let us consider the interaction $W_{ab}$ between the valence
nucleon $b$ and the even-even core containing paired nucleons $a$.
Under the assumption that the collective octupole excitation does
not involve spins of the individual core nucleons, we will keep
only the terms that do not contain spin operators
\mbox{\boldmath$\sigma$}$_{a}$ being related to the spin current
of the external nucleon. Using the contact nature of the
interaction (\ref{wfull}) and taking the matrix element over the
state $\psi_{b}$ of the external nucleon $b$ we can present the
operator acting onto the core nucleons in the following form:
\begin{equation}
\hat{W}_{a}=-\frac{G\eta_{ba}}{\sqrt{2}}\frac{1}{2m} \left(
\mbox{\boldmath$\nabla$}_{a}\cdot \psi_b^{\dagger}({\bf
r}_a)\mbox{\boldmath$\sigma$}_{b}\psi_b({\bf r}_a) \right).
\label{wa}
\end{equation}

The operator $\hat{W}_{a}$ contains the octupole component
proportional to the operator $\hat{O}_{\rm intr}=r^3 Y_{30}$ and
responsible for the collective excitation. To extract this
component, we project out the octupole operator in the intrinsic
frame,
 \begin{equation}
\hat{W}_{a}=C_{a}\hat{O}_{\rm intr} +...\,,
\label{woct}
\end{equation}
where the amplitude of this component, the contribution of a core
nucleon $a$ to the octupole mode, is
 \begin{equation}
C_{a}= \frac{\langle\hat{O}_{\rm intr}|\hat{W}_{a}\rangle}
 {\langle\hat{O}_{\rm intr}|\hat{O}_{\rm intr}\rangle_R}.
\label{coct}
\end{equation}
Here we omitted the electric charge $e$ from the definition of the
octupole operator $\hat{O}_{\rm intr}$ since we assume that it
acts both on protons and neutrons (an isoscalar octupole mode). In
the collective transition to the excited octupole state the matrix
element of the octupole operator $\hat{O}_{\rm intr}$ is enhanced.
Therefore, we assume that we can neglect all terms in the
expansion (\ref{woct}) except for the one written down explicitly.

For calculating the projection constant $C$ we assume that the
amplitude of the octupole vibrations is small compared to the
nuclear radius so that the operator $\hat{O}_{\rm intr}=r^3
Y_{30}$ is acting effectively only within the nucleus,
$\hat{O}_{\rm intr}=0$ for $r>R +\delta$ where $\delta \ll R$ is
some small distance. This allows us to introduce the normalized
``octupole'' state ${\rm const} \cdot r^3 Y_{30}$ and consistently
define the projection procedure. In this case
 \begin{equation}
\langle\hat{O}_{\rm intr}|\hat{O}_{\rm intr}\rangle_R=\int_0^R r^6
Y_{30}^2 d^3r=R^9/9, \label{norm}
\end{equation}
and
\begin{equation}
\langle\hat{O}_{\rm intr}|\hat{W}_{a}\rangle=
\frac{G\eta_{ba}}{\sqrt{2}}\frac{1}{2m} \int \Psi_b^+({\bf
r})\mbox{\boldmath$\sigma$}\Psi_b({\bf r})\cdot
 \left( \mbox{\boldmath$\nabla$}r^3 Y_{30} \right) d^3 r=
 \frac{9\sqrt{7}}{40 \sqrt{2\pi}} \frac{G\eta_{ba}}{m} R^2 F.     \label{wo}
\end{equation}
Here we introduced the expectation value over the orbital of the
external nucleon,
\begin{equation}
F=\frac{\langle r^2 ( 5 n_z^2 \sigma_z -\sigma_z + 2 n_z
(\mbox{\boldmath$\sigma$}\cdot {\bf n}))\rangle}{(3/5) R^2},
\label{F}
\end{equation}
where ${\bf n}={\bf r}/r$ is the unit vector, $n_z=\cos\theta$.
Note that $\langle r^2\rangle \approx 3 R^2/5$, therefore $F \sim
1$. In the case of $^{223,225}$Ra, assuming that the unpaired
neutron is in the (asymptotic) Nilsson orbital $1/2[631]$ with
$\sigma_z$=+1 or $-1$, we would get $|F| \approx 0.3$. For a pure
$ls$-state $|lm \pm\rangle$ with $m_{l}=m, \, s_{z}=\pm (1/2)$,
this factor is given by
\begin{equation}
\langle lm\pm|F|lm\pm\rangle=\pm
2\frac{l(l+1)-3m^{2}}{(2l+3)(2l-1)} \,\frac{\langle
r^{2}\rangle}{(3/5)R^{2}}.                      \label{n0}
\end{equation}
Here the angular factor in front of $\langle
r^{2}\rangle/(3R^{2}/5)$ changes for even $l$ from 4/7 to 1/2 if
$m=0$ and from -2/7 to -1/2 if $m=1$. For the realistic Nilsson
wave functions with the quadrupole deformation around 0.15 or 0.2,
the spin-orbital coupling is important, and the wave function of
the valence neutron is a superposition of several spherical
orbitals $|l, m\mp 1/2,\pm 1/2\rangle$. For example, the orbital
$1/2[631]$ has 7 main components $|Nlms_{z}\rangle$ of comparable
magnitude, where $N$ is the major oscillator quantum number,
\begin{equation}
|1/2[631]\rangle\approx \sum_{l=0,2,4,6}a_{l}|6l0\,1/2\rangle+
\sum_{l=2,4,6}b_{l}|6l1-1/2\rangle.
\label{n1}
\end{equation}
Although the last item in the numerator of the expression
(\ref{F}) contains the matrix elements with $\Delta s_{z}=\pm 1$
and is sensitive to the interference of the $\sigma_{z}=\pm$
components that might be constructive or destructive, the quantity
$|F|$ is typically between 0.1 and 0.6. For the specific values of
the parameters $a_{l}$ and $b_{l}$ corresponding to the Nillson
wave function 1/2[631] at $\beta_{2}=0.21$, we come to $F=-0.32$, very
close to the asymptotic estimate.

Combining the equations above we obtain the effective ``octupole''
${\cal P},{\cal T}$-odd operator
 \begin{equation}
\hat{W}_{{\rm oct}}= \frac{\langle\hat{O}_{\rm
intr}|\hat{W}_{a}\rangle}
 {\langle\hat{O}_{\rm intr}|\hat{O}_{\rm intr}\rangle_R} \hat{O}_{\rm intr}=
 \frac{81}{40}\sqrt{\frac{7}{2\pi}}\, \frac{G\eta_{ba}F}{m R^7}\,
\hat{O}_{\rm intr}.
\label{weakoct}
\end{equation}
Using Eq. (\ref{b3}) for the proton induced transition and summing
over proton and neutron contributions to the octupole excitation
$|{\rm oct}\rangle$, we obtain the expression for the weak matrix
element in the body-fixed frame,
\begin{equation}
\langle {\rm g.s.}|W|{\rm oct}\rangle=
\frac{81}{40}\sqrt{\frac{7}{2\pi}}\,
 \frac{G\eta_{b}F}{m R^7} AR^{3}\frac{3}{4\pi}\beta_{3}^{{\rm rms}},
\label{woctupole}
\end{equation}
where $\eta_b=\frac{Z}{A} \eta_{bp} + \frac{N}{A} \eta_{bn}$,
$b=n$ for the external neutron or $b=p$ for the external proton.

The transition from the body-fixed  frame to the laboratory frame
for the vector Schiff moment gives, as in Eq. (\ref{nz}), an extra
factor $KM/I(I+1)$ that, for the ground nuclear state, $K=M=I$, is
equal to $I/(I+1)$. Collecting all factors we obtain the Schiff
moment due to the octupole collective vibrations:
\begin{equation}
{\bf S}=\frac{2I}{I+1}\frac{\langle {\rm g.s.}|W|{\rm oct}\rangle
\langle {\rm oct}|{\bf S}|{\rm g.s.}\rangle}{\Delta E},
                                                               \label{Svect}
\end{equation}
\begin{equation}
S=0.025 F \,\frac{I}{I+1} ~e
\beta_{2}(\beta_{3}^{{\rm rms}})^{2}ZA^{2/3} \frac{\eta_b G}{mr_0 \Delta
E},      \label{schiffsoft}
\end{equation}
where $\Delta E=E_{{\rm g.s.}}-E_{{\rm oct}}$, or, in appropriate
numerical units,
\begin{equation}
S=\frac{Z}{94}\,\left(\frac{A}{230}\right)^{2/3}\frac{\beta_{2}}{0.2}\,
\left(\frac{\beta_{3}^{{\rm rms}}}{0.1}\right)^{2}\frac{(5/3)\langle
r^{2}\rangle} {R^{2}}\,\frac{50\,{\rm keV}}{\Delta
E}\frac{I}{I+1}\,F(5650\times 10^{-8}) e\eta\,{\rm fm}^{3}.
\label{num}
\end{equation}
The ratio of this dynamical (vibrational) octupole contribution
$S_{{\rm soft}}$ to the static octupole contribution $S_{{\rm
stat}}$ in the deformed case (\ref{snum}) is equal to
\begin{equation}
\frac{S_{\rm soft}}{S_{\rm stat}}=2.5 F\frac{(\beta_{3}^{{\rm rms}})^2}
{(\beta_{3}^{{\rm stat}})^2} \approx \frac{(\beta_{3}^{{\rm rms}})^2}
{(\beta_{3}^{{\rm stat}})^2} . \label{schiffratio}
\end{equation}
This means that the dynamical deformation due to low-frequency
octupole vibrations produces nearly the same effect as the static
octupole deformation.

When doing numerical estimates we may assume
$\beta_{3}^{{\rm rms}}=\beta_{3}^{{\rm stat}}=0.1$ \cite{AFS,SAF,hayes}.
This gives numerical values (in the units of $10^{-8}\eta e\cdot {\rm
fm}^{3}$) for the Schiff moments of $^{223}$Ra and $^{225}$Ra,
$S=700$ and 400, correspondingly;  $S=1000$ \cite{Rn} for
$^{223}$Rn; and in the interval 100 - 1000 for $^{223}$Fr,
$^{225}$Ac and other nuclei with the low octupole mode, $\Delta E<100$ 
keV, i.e. close to the values
\cite{AFS,SAF} obtained in the numerical calculations for the
static octupole deformation. However, our value for $^{239}$Pu,
where $\Delta E=470\, $keV, is $S \sim$ 80,
or even smaller if we take into account that in the ground state
of a harmonic oscillator the vibrational amplitude is proportional
to $\omega^{-1/2}$ so that $\langle \beta^2\rangle \propto
1/\omega$ as it is approximately the case for both quadrupole and
octupole low-lying vibrations in nuclei \cite{Ram,Met}. This
result is much smaller than the estimates $S$(Pu)=300$ \cdot
S$(Hg) $\approx 420 $ presented in Ref. \cite{hayes}.

The work \cite{hayes} also claims that the contribution of the
octupole vibration to the $^{199}$Hg Schiff moment is about half
of its shell model value. Strictly speaking, our formula is not
applicable for this case since $^{199}$Hg, probably, has no
quadrupole deformation. However, for an estimate we can take
 $\beta_2=0.1$ assuming that a similar contribution may appear due to an
interaction with soft quadrupole vibrations. Now we can make a
comparison between $^{225}$Ra and $^{199}$Hg which both have
$I=1/2$. The octupole transition energy in $^{199}$Hg is 3000 keV
\cite{hayes}. Therefore, in the case of $^{199}$Hg we lose one-two
orders of magnitude due to the octupole deformation [if
$\beta_3^2\propto(\Delta E)^{-1}$] and one-two
orders of magnitude due to the energy denominator that is sixty
times greater than in $^{223,225}$Ra. This gives the octupole
contribution to the $^{199}$Hg Schiff moment about $400 \cdot
0.001 \sim 0.4$. The shell-model contribution is $-1.4$. This
estimate shows that the octupole contribution should not produce
any dramatic changes in the $^{199}$Hg Schiff moment (in fact,
this conclusion agrees with that of Ref. \cite{hayes}).

\subsection{Particle excitation}

In the previous consideration the orbital of the valence nucleon
in the mean field of a normal quadrupole deformation was fixed. In
order to estimate the contribution of the abovementioned second
mechanism related to the mixing of single-particle orbitals by the
octupole mode, we assume that the octupole component of the strong
nuclear Hamiltonian can be estimated in terms of octupole-octupole
forces \cite{bm} as $\beta_{3}V_{3}$, where $\beta_{3}$ acts on
the core variables exciting octupole vibrations while
$V_{3}=f(r)Y_{30}$ is the perturbation acting on the valence
nucleons. For the small amplitude $\beta_{3}$, $f(r)$ is
proportional to the derivative of the mean nuclear potential. As
pointed out in Ref. \cite{hayes}, this interaction mixes the
ground state $|\lambda;0\rangle$ of an odd-$A$ nucleus (a valence
particle on a single-particle orbital $\lambda$ and the even-even
core in the ground state) with the excited states $|\nu;1\rangle$
that contain a particle in an orbital $\nu$ of opposite parity and
a phonon of the octupole mode. To the first order in this mixing,
the new ground state is
\begin{equation}
|\tilde{0}\rangle
=|\lambda;0\rangle+\sum_{\nu}c_{\nu}|\nu,1\rangle. \label{n1a}
\end{equation}
The admixture coefficients are given by
\begin{equation}
c_{\nu}=\frac{\langle 1|\beta_{3}|0\rangle (V_{3})_{\nu\lambda}}
{\epsilon_{\lambda}-\epsilon_{\nu}-\hbar\omega},
\label{n2}
\end{equation}
where $\epsilon_{\lambda}$ and $\epsilon_{\nu}$ are
single-particle energies whereas $\hbar\omega$ is energy of the
octupole core excitation. Similarly, the unperturbed one-phonon
state $|\lambda;1\rangle$ also acquires new components being
converted into
\begin{equation}
|\tilde{1}\rangle=|\lambda;1\rangle+\sum_{\nu}c_{\nu}'|\nu;0\rangle,
\label{n3}
\end{equation}
where
\begin{equation}
c_{\nu}'=\frac{\langle 0|\beta_{3}|1\rangle (V_{3})_{\nu\lambda}}
{\epsilon_{\lambda}-\epsilon_{\nu}+\hbar\omega},
\label{n4}
\end{equation}
and we omit the two-phonon admixtures irrelevant for our purpose.

The states $|\tilde{0}\rangle$ and $|\tilde{1}\rangle$ can be
mixed by the effective ${\cal P},{\cal T}$-odd potential $W$
averaged, in distinction to the previous case of Eq. (\ref{wa}),
over the core nucleons. For a conventional definition of angular
wave functions, the matrix elements of $\beta_{3}, \,V_{3}$ and
$W$ are real, and the mixing matrix element is
\begin{equation}
\langle \tilde{0}|W|\tilde{1}\rangle=\langle
1|\beta_{3}|0\rangle\sum_{\nu}
\frac{2(\epsilon_{\lambda}-\epsilon_{\nu})}{(\epsilon_{\lambda}-
\epsilon_{\nu})^{2}+(\hbar\omega)^{2}}W_{\lambda\nu}(V_{3})_{\nu\lambda}
\equiv \langle 1|\beta_{3}|0\rangle\langle\tilde{0}|W_{0}|\tilde{0}\rangle.
                                                            \label{n5}
\end{equation}
where the single-particle matrix elements can be renormalized by
the pairing correlations. In the limit of a soft octupole mode,
when $(\hbar\omega)^{2}$ is small compared to the typical
single-particle energy intervals $(\epsilon_{\lambda}
-\epsilon_{\nu})^{2}$ between the orbitals of opposite parity, the
latter expression coincides with the result for the static
octupole deformation found in Ref. \cite{SAF}. This leads to a
conclusion that the second mechanism in the presence of the soft
octupole mode provides the Schiff moment
\begin{equation}
S_{2}\approx \frac{2I}{I+1}\,\frac{|\langle
1|\beta_{3}|0\rangle|^{2}} {\Delta
E}\,\frac{9}{20\pi\sqrt{35}}\,eZR^{3}\beta_{2}\langle\tilde{0}|W_{0}|
\tilde{0}\rangle                                    \label{S2}
\end{equation}
that can be described by Eq. (\ref{snum}), derived for the static
octupole contribution, with the replacement of
$(\beta_{3}^{{\rm stat}})^{2}$ by $(\beta_{3}^{{\rm rms}})^{2}=|\langle
1|\beta_{3}|0\rangle|^{2}$. It is easy to explain this result. 
For an adiabatically slow
vibrational motion, we can calculate the weak and Schiff moment
matrix elements at a fixed value of $\beta_{3}$ and then average
over $\beta_{3}$. This conclusion holds for the first mechanism as
well. Indeed, keeping the linear in $\beta_{3}$ terms, we can
present the effective ${\cal P},{\cal T}$-odd potential $W$
averaged over the core nucleons as  $W=W_{0}+\beta_{3}W_{3}$. This
leads to the following contribution of the octupole weak field
\begin{equation}
S_{2}\approx \frac{2I}{I+1}\,\frac{|\langle
1|\beta_{3}|0\rangle|^{2}} {\Delta
E}\,\frac{9}{20\pi\sqrt{35}}\,eZR^{3}\beta_{2}\langle 0 |W_3|
0\rangle                                             \label{S2a}
\end{equation}
Here the expectation value is taken over the ground state
$\lambda$ of the valence nucleon ($\langle 0 |W_3| 0\rangle =
 \langle\lambda |W_3|\lambda\rangle$). 
Again, the replacement of $(\beta_{3}^{{\rm rms}})^{2}=|\langle 1|
\beta_{3}|0\rangle|^{2}$ by $(\beta_{3}^{{\rm stat}})^{2}$ reduces the
dynamical problem to the static one. The final result for the
first mechanism contribution was given in eq. (\ref{schiffsoft}).

\section{From nuclear Schiff moment to atomic EDM}

Finally, we should calculate the atomic  electric dipole moments
induced by the nuclear Schiff moments.

The atomic EDM is generated by the ${\cal P}, {\cal T}$-odd part
of the nuclear electrostatic potential $\varphi({\bf r})$. The
potential produced by the point-like Schiff moment is usually
presented in the form \cite{FKS84} proportional to the gradient of
the delta-function at the origin,
\begin{equation}
\varphi ({\bf r})=4\pi {\bf S}\cdot \mbox{\boldmath$\nabla$}\delta
({\bf r}).
                                                            \label{olddef}
\end{equation}
The natural generalization of the Schiff moment potential for a
finite-size nucleus is \cite{ginges}
\begin{equation}
\varphi ({\bf r})=-3 ({\bf S}\cdot {\bf r})\,\frac{n(r)}{B}\ ,
\label{phigen}
\end{equation}
where $B=\int n(r)r^{4}dr\approx R^{5}/5$, $R$ is the nuclear
radius, and $n(r)$ is a smooth function which is $1$ for
$r<R-\delta$ and $0$ for $r>R+\delta$ while $\delta\ll R$ is some
small distance; $n(r)$ can be taken as proportional to the nuclear
density. The potential (\ref{phigen}) corresponds to a constant
electric field \mbox{\boldmath${\cal E}$} inside the nucleus that
is directed along the nuclear spin, \mbox{\boldmath${\cal E}$}
$\propto {\bf I}$. The interaction $-e\varphi$ mixes electron
orbitals of opposite parity and produces EDMs in atoms.

In Ref. \cite{dzuba} we have performed atomic calculations of the
coefficients $k$ that define the atomic EDM $d$ in terms of the
Schiff moments,
\begin{equation}
d=k \cdot 10^{-17}\cdot \left(\frac{S}{e \cdot {\rm
fm}^3}\right)\, e \cdot {\rm cm} .
\label{d}
\end{equation}
The factors $k$ rapidly grow with the nuclear charge that leads to
an additional enhancement in Ra, $k=-8.5$,  and Rn, $k=3.3$, in
comparison with the lighter electronic analogues, Hg, $k=-2.8$,
and Xe, $k=0.38$.

We have estimated above the individual contributions of two
possible mechanisms for the enhancement of original ${\cal P},
{\cal T}$- odd interactions $W$ by nuclear soft octupole degrees
of freedom. The virtual core excitation and the mixing of
single-particle states work in fact together and lead to the
resulting enhancements of the same order of magnitude. In
principle they may involve different parts of the interaction $W$.
The detailed calculation of the total result is subject to many
uncertainties both on the fundamental and nuclear structure
levels; this should be studied separately. However, there is no
physical reason for expecting their strong cancellation. As we
have seen in the consideration of the core excitation mechanism,
the exact magnitude of the resulting Schiff moment is quite
sensitive to the details of the valence single-particle wave
function in the presence of spin-orbit coupling, whereas the
second mechanism is determined mainly by the intruder orbits of
opposite parity. With no accidental cancellation, the resulting
Schiff moment will be of the same order of magnitude as each of
the contributions. Taking some conservative ``minimal'' values of
the Schiff moments for the different methods of calculations
(static analytical, static numerical, soft analytical) we obtain
the values of the atomic electric dipole moments given in the
Table.
\begin{center}
\begin{tabular}{c|c|c}
Nucleus & $S,\,10^{-8}\eta_{n}e\cdot{\rm fm}^{3}$ &
$d,\,10^{-25}e\cdot{\rm cm}
\cdot\eta_{n}$\\
\hline
$^{225}$Ra & 300  & 2500\\
$^{223}$Ra & 400  & 3400\\
$^{223}$Rn & 1000 & 3300
\end{tabular}
\label{Table}
\end{center}

\noindent
This may be compared with typical values of the induced dipole
moment predicted for spherical nuclei, $d=4$ and $d=0.7$ for
$^{199}$Hg and $^{129}$Xe, respectively.

To conclude, we made analytical estimates for the nuclear Schiff
moment in nuclei known as soft with respect to the octupole
excitation mode. We found a strong enhancement of the average
magnitude close to what was found for nuclei having static
octupole deformation. There are several factors which contribute
to this enhancement. Mainly, they are small energy interval
between the opposite parity states and large amplitude of vibrations in the 
case of the soft octupole mode. Effectively, the enhancement factor can be 
presented as const$\times (E/\hbar \omega)^2$, where $E$ is the intershell 
distance in spherical nuclei, $\omega$ is the frequency of the octupole mode 
and const$\simeq 0.1$.
 The value of the Schiff moment is typically
$10^{2}\div 10^{3}$ in units of $10^{-8}\eta e\cdot{\rm fm}^{3}$;
it leads to the prediction of the enhancement of the atomic EDM on
the level of
 $ 10^{3}$ as compared to spherical nuclei.

A related idea should be explored in the future: it is known that
some nuclei are soft with respect to both quadrupole and octupole
modes, see for example recent predictions for radioactive nuclei
along the $N=Z$ line \cite{Kaneko}. The light isotopes of Rn and
Ra are spherical but with a soft quadrupole mode and therefore
large amplitude of quadrupole vibrations. The spectra of these
nuclei display long quasivibrational bands based on the ground
state and on the octupole phonon, with positive and negative
parity, respectively. These bands are connected via low-energy
electric dipole transitions. This situation seems to
be favorable for the enhancement of ${\cal P},{\cal T}$-odd effects.\\
\\
{\small The authors are thankful to the Institute of Nuclear
Theory, University of Washington, where this work was started, for
hospitality and support. The discussions with A. Dieperink, O.
Scholten and J. Engel are gratefully acknowledged. V.Z.
acknowledges support from the NSF grant PHY-0070911. V.F.
acknowledges support from the Australian Research Council. He is
also grateful to the Institute for Theoretical Atomic and Molecular Physics,
Harvard-Smithsonian Center for Astrophysics, where part of this
work has been done.}


\begin{thebibliography}{99}
\bibitem{comment} Note that the numerical calculation in
\cite{SAF} includes effects of the octupole deformation in  both
the strong and weak potentials - see Eqs. (30)-(31) in \cite{SAF}.
This is important since these two contributions tend to cancel
each other. A comment in \cite{hayes} that the second contribution
was neglected in \cite{SAF} is erroneous.
\bibitem{Rn} The energy denominator for $^{223}$Rn is not known
experimentally and the particle-rotor model \cite{leander} does
not give a unique prediction. Therefore we take the energy
splitting 50 keV, similar to $^{223}$Ra and $^{225}$Ra. The
calculated value of 37 keV \cite{SAF} is even smaller.
\end{thebibliography}
\end{document}